\documentclass[namedreferences]{kluwer}

\usepackage{epsfig}
\usepackage{times}
\usepackage{mathptm}

\input epsf
\def\eps@scaling{.95}
\def\epsscale#1{\gdef\eps@scaling{#1}}
\def\plotone#1{\centering \leavevmode
    \epsfxsize=\eps@scaling\columnwidth \epsfbox{#1}}
\def\plottwo#1#2{\centering \leavevmode
    \epsfxsize=.45\columnwidth \epsfbox{#1} \hfil
    \epsfxsize=.45\columnwidth \epsfbox{#2}}

\newcommand{\um}{\,\hbox{$\mu$m}}
\def\deg{\nobreak{$^\circ$}}
\def\arcsec{\nobreak{$''$}}
\newcommand{\mv}{\hbox{m$_{\rm V}$}}
\newcommand{\km}{\,\hbox{km}}
\def\spose#1{\hbox to 0pt{#1\hss}}
\def\simlt{\mathrel{\spose{\lower 3pt\hbox{$\mathchar"218$}}
     \raise 2.0pt\hbox{$\mathchar"13C$}}}
\def\simgt{\mathrel{\spose{\lower 3pt\hbox{$\mathchar"218$}}
     \raise 2.0pt\hbox{$\mathchar"13E$}}}
\newcommand{\hiiregion}{\hbox{H\,{\sc ii}~region}}

\begin{document}

\begin{article}

%
%

\begin{opening}

\title{The ALFA Laser Guide Star: Operation \& Results}

\author{R. \surname{Davies}\email{davies@mpe.mpg.de}}
\institute{Max-Planck-Institut f\"ur extraterrestrische Physik, Garching}
\author{A. \surname{Eckart}}
\author{W. \surname{Hackenberg}}
\author{T. \surname{Ott}}
\author{D. \surname{Butler}}
\institute{Max-Planck-Institut f\"ur extraterrestrische Physik, Garching}
\author{M. \surname{Kasper}}
\institute{Max-Planck-Institut f\"ur Astronomie, Heidelberg}
\author{A. \surname{Quirrenbach}}
\institute{University of California, San Diego}


%
\begin{abstract}
The importance of laser guide stars to the practical usefulness of
adaptive optics cannot be understated, and yet there are very
few working systems.
This contribution discusses the current status of the ALFA
laser guide star, with regard to the particular difficulties encountered
while observing as well as both the expected performance and that so far
achieved from scientific observations.
A description is given of a number of experiments involving ALFA which
aim to determine the atmospheric turbulence and sodium layer
charateristics, and demonstrate the possibility for tilt recovery from
laser guide stars.
\end{abstract}



\end{opening}

%
%

\section{Introduction} 
\label{sect:intro}

The resolution and sensitivity that should be achievable with
the current generation of 8-m class telescopes which are either on-line
or being built, has the potential to revolutionise astrophysics.
But in order to reap these benefits it is essential to use adaptive
optics techniques, and with the requirement for maximal sky coverage
comes the mandatory use of laser guide stars.

Over recent years there has been an increasing effort put into the
development of high-order AO instrumentation, as well as observational
programmes: 
the number of refereed papers based on AO has been increasing dramatically
(27 in 1997 compared to only 18 in the entire 4 years prior to that,
\opencite{rid98}), and recently an entire meeting was
devoted to scientific results using adaptive optics systems 
({\em Astronomy with Adaptive Optics}, ESO 1998).
However, this also highlighted the fact that nearly all
observations to date have used natural guide stars
almost exclusively as the wavefront reference.
This has resulted in the majority of results being concerned with
stellar systems, while relatively few have been on extragalactic sources.
The bias can be explained by the lack of suitable reference stars, since
the most commonly observed extragalactic objects lie out of the
galactic plane, while most stellar sources lie in the galactic plane
where the bright reference stars can be found;
and with the exception of a few of the brightest AGN, the nuclei of other
galaxies are too faint for wavefront sensing with current technology.
Only partial correction, and that under exceptional conditions, can be
achieved while correcting on even the brightest starburst hotspots
\cite{kna99}.

Clearly there is a need for further work to develop high quality
artifical guide stars, but technical difficulties have slowed progress.
ALFA ({\bf A}daptive optics with a {\bf L}aser {\bf F}or
{\bf A}stronomy), is addressing these issues
and has made significant progress in this area.
It is installed on the 3.5-m telescope at Calar Alto, and is now 
the first laser guide star (LGS) adaptive optics facility to be
available to guest observers.
In this contribution, by discussing the operation of and
observational results from the ALFA LGS, we attempt to provide the
reader with an understanding that will enable him/her to answer the
question ``How useful are laser guide stars at the present time?''.

Section~\ref{sec:ov} introduces ALFA with a brief overview of the
developmental milestones.
The observational practice is then outlined in Section~\ref{sec:operate},
followed by a discussion of the limiting features of the LGS.
Section~\ref{sec:tmr} briefly mentions a number of experiments that are
being carried out in conjunction with the European Laser Guide Star
Network in order to better understand, quantify, and improve the
performance of LGS.
This must be done before they can be usefully
employed on 8-m class telescopes.
Lastly in Section~\ref{sec:res} we present some observational results
that indicate the current performance of the system.

\section{Overview}
\label{sec:ov}

The performance goal of ALFA was to reach 50\% Strehl at
2.2\um\ under typical seeing conditions (0.9\arcsec) with good sky
coverage.
Due to the technical challenge inherent in building both an adaptive
optics bench and a laser system, these were constructed separately and
are described below.

The AO system is based on a Shack-Hartmann sensor with
several interchangeable lenslet arrays, and is described in detail in
\inlinecite{hip98}, \inlinecite{wir98},
and Kasper et al. (this issue).
It is optimised for partial correction using a LGS,
with the aim of typically sampling 18 hexagonal subapertures at a frame
rate of 100\,Hz.
In this configuration, the centroids of the spots in the sensor are
used to determine the 
coefficients for 21 Zernike or 24 Karhunen-Loeve modes, and
these are then applied to a 97-actuator mirror.
The tip and tilt components must be measured using a separate tracker
camera since it is not possible to distinguish the up-link motion of
the laser beam from the down-link tipt-tilt motion.

The optical bench for the laser is installed in the coud\'e lab, where
an Ar$^+$ laser pumps a continuous-wave dye ring laser with a single
line output power of 3.8\,W.
The output beam is fed along the coud\'e path
and directed into a launch telescope, resulting in a sodium beacon with
a magnitude of about m$_{\rm V} = $9--10 at zenith.
Further details of the lasers, beam relay, diagnostic tools, and safety
issues are given by Rabien et al. (this issue) and Ott et al. (this issue).

Since Sept 1997, when the loop was locked on the laser for the first
time, development of the LGS and AO has progressed in tandem and
been fairly rapid.
In December 1997, the LGS was used to enhance image resolution,
sampling 6 subapertures at a frame rate of 60\,Hz, and correcting 7
modes (plus tip-tilt).
During these observations the binary stars in BD+31\deg643 were clearly
resolved.
In March 1998, the laser guide star was seen on the wavefront sensor
using 18 subapertures at a frame rate of 100\,Hz.
A summary of the status and performance at this time can be found in
\inlinecite{dav98d}.
Since then significant improvements to the hardware
have enhanced the performance considerably.
Although poor weather has hampered testing of the upgraded system,
during an August observing run K-band Strehl ratios higher than 60\%
were measured on bright stars, and even on V-11\,mag stars Strehl
ratios of 10--15\% within a radius of 50$''$ were achieved.
Also, for the first time we successfully used the laser guide star to
improve the image quality of another galaxy, UCG\,1347 in the Abell\,262
cluster.
Our most recent success, in June 1999, is to achieve a Strehl ratio
greater than 20\% in the K-band.

\section{Operation}
\label{sec:operate}

\subsection{Observing with a LGS}

The entire operating procedure for running the laser and
observing are described in 
`{\it The ALFA Laser Guide Star System: A User's Guide}' 
which is available on-line from the ALFA pages at
\newline{\tt http://www.mpe.mpg.de/www\_ir/ALFA/ALFAindex.html}.
A few of the salient points concerned with observational issues are
discussed below.

There is a set procedure for optimising the LGS at the beginning of
every night which is currently semi-automated.
The rayleigh cone and LGS show up clearly on the TV finder and this
allows correct positioning, a step which must always
be executed manually.
The aim is to locate the LGS on the wavefront sensor (with a field of
view of only a few arcsec) for fine tuning and focussing, a
3-stage process requiring adjustment of laser frequency,
launch telescope and WFS.
Typically the frequency does not need adjusting and it is sufficient to
tune it to the zero velocity component observed in a Na-cell in the lab.
An automated algorithm then takes care of the focussing, and has
increased the observing efficiency by a alrge margin.
The principle of the algorithm is similar to that used on many
telescopes:
measure the PSF at a number of focus positions which straddle the
optimal one, fit a curve through the points, and read off the best
position.
The only subtle part is how best to measure the PSF: brightest pixel,
encircled energy, FWHM, etc.
for launch telescope and WFS focussing, the ratio of the flux in
small and large apertures should be a good measure of how concentrated
the energy is.
A single iteration of the algorithm for each focus required appears to
be sufficient.
Once the LGS is as compact and bright as possible the wavefront sensor
can be calibrated using a reference fibre.
However, because the height of the Na layer is not known exactly, the
focus position at which the reference fibre must be placed is
also uncertain.
Thus another iteration of the algorithm is used to focus the fibre on
the WFS, by moving the fibre along the optical axis while keeping the
WFS itself fixed.

The bright Rayleigh cone which extends to a height of $\sim$25\,km 
is separated from the LGS by about 17$''$ at
zenith and is rejected from the wavefront sensor simply by using a
10$''$ aperture.
Although the separation lessens at large zenith angles, this is more
than sufficient since even at 60\deg\ it is still more than 8$''$ so the
light pollution in the sensor is low.
Recent observations have shown that there is still scattered light in
the WFS aperture, although at a relatively faint level, which may be
reflected from inside the dome.
A smaller field-stop will be used to reduce this as much as possible.

As the telescope tracks, the distance to the sodium layer changes and so
periodic tuning, focussing, and calibration must be carried out.
At present we have not made long enough exposures to need to do this
more frequently than every time we slew to a new object.
Currently we do not expect to need to do this more than every
1--2~hours, but once the LGS spot is optimised (see
Section~\ref{sec:size}) more frequent adjustments may become necessary.

\subsection{LGS Properties}
\label{sec:prop}

In March~1998 the LGS was seen on the wavefront sensor through
18 subapertures at a frame rate of 100\,Hz, as reported in
\inlinecite{dav98d}.
However, this is still  difficult to achieve since 
the LGS is rather more extended and fainter than a star of similar
V-magnitude.
Initially, the LGS brightness was measured in magnitudes for easy
comparison to stars, and this has stuck even though it is misleading
--- 
due to the broad sensitivity range of the wavefront sensor CCD (the
quantum efficiency remains above 50\% over 0.49--0.84\um) which allows
it to detect emission in the V- and R-bands as well as some of the
I-band.
A more natural unit would be the photon number detection rate, which is
appropriate for the narrow line width of the LGS and is also
independent of a star's spectral type.
However, this has not been adopted as it is extremely inconvenient when
searching (magnitude based) catalogues for possible guide stars.
It should therefore be borne in mind that the number of photons
detected from a \mv=9.5 LGS will be the same as those from a
\mv=10.9 A0 star.
Additionally, more photons are detected from redder stars: the above
example could be extended to include a K0 star reddened by 1 mag, which
need only have \mv=11.5.

The size and brightness are the two observables that must be optimised,
and they are considered below before we show some simulations of the 
performance that we should expect from the LGS.

\subsubsection{Size \& Shape}
\label{sec:size}

The angular size of the LGS needs to be minimised so that
it is more intense and the centroid better defined, requiring the
projected beam to be focussed on the Na layer from a launch
telescope.
Two conflicting effects determine the diameter $D$ required:
1) in the absence of turbulence, simple diffraction theory says that the
theoretical minimum spot size is smaller when a wider exit-beam
diameter is used ($\theta_{\rm min} = \lambda/D$);
2) but the presence of atmospheric turbulence means that if the exit-beam
exceeds 2--3 times the coherence length $r_0$, the spot will break up
into speckles, increasing the effective size to $\theta = \lambda$/$r_0$.
In order to cope with variable conditions on different nights there is
a pre-expander on the optical bench which increases the beam diameter to
1.5--3.0\,cm, controlling the width of the beam exiting the
launch telescope within the range 24--49\,cm, suitable for seeing of
0.8--1.5\arcsec\ ($r_0 = 8$--15\,cm at 589\,nm).
The single (Gaussian) TEM$_{00}$ mode quality of the beam also helps to
concentrate most of the power in a small region.

Up until mid-1999, the LGS had a typical FWHM of $\sim$3\arcsec, the
cause of which was uncertain.
Originally the situation was worse, and considerable turbulence arose
between the coud\'e lab and the telescope.
To combat it the entire beam path was enclosed in pipe sections, a
scheme that produced a noticeable improvement and additionally acted as
a safety precaution.
To measure the distortion, 
a Shack-Hartmann sensor was installed in the diagnostics bench
below the launch telescope (\opencite{rab99}, also this issue).
Because it does not operate in a photon-starved environment no special
equipment is required, and the main components are an old lenslet array
from ALFA's AO bench and a 30\,Hz frame-rate camera.
The results indicated an rms wavefront error of only
0.1$\lambda$ (at the laser wavelength of 589\,nm), which is much less
than that required to explain the LGS size.
The lack of a calibration fibre meant that this WFS could only measure
dynamic aberrations, so a Shearing interferometer was used to
study the static aberrations.
It suggested that a few optical elements in the beam path (eg
the glass plate sealing the end of the pipe from the coud\'e lab) were
to some extent culprits of the poor beam quality.
It finally became clear, by imaging stars through the launch telescope,
that this was the source of most aberrations.
Since installation it had become misaligned, and additionally the
primary mirror was being slightly distorted by the clamps holding it in
place.
A new launch telescope is been constructed which should remove the
problem entirely; but by careful correction of the old one we are able
to reduce the spot to 1.7\arcsec, even using a launch beam only
10\,cm wide and in seeing of 1.2\arcsec.

\inlinecite{rou94} gives the phase measurement error
$\sigma^2$~(rad$^2$) associated with signal photon noise in a
Shack-Hartmann sensor as 
\[
\sigma^2 \ \ = \ \ 
			\frac{\pi^2}{2 \ n_{\rm ph}}
			\left( \frac{\theta_{\rm lgs} \ d}
							{2 \times10^{-4} \ \lambda} \right)^2
\]
where $n_{\rm ph}$ is the number of photons of wavelength
$\lambda$~(nm) detected in one subaperture of diameter $d$~(m) in one
sampling time, and $\theta_{\rm lgs}$~(arcsec) is the angular diameter
of the LGS.
Clearly, if we could reduce the spot size to
only 1\arcsec\ then the photon noise phase error would be reduced by an
order of magnitude.
Only then would elongation of the spot due to off-axis
projection become an issue, although
given the axial separation of 2.9\,m between the telescopes, the
elongation will be in the range 0.23--0.95\arcsec\ (near and far-sides
of the telescope primary) for an 8\km\ thick sodium layer 
at a mean altitude of 90\km.
Thus for a 1\arcsec\ LGS, the aspect ratio would be less than 1:1.4.
With our current set-up and LGS spot size this is not a problem, but
more detailed modelling of the LGS elongation in different subapertures
both for ALFA and for the VLT can be found in Viard et al. (this
issue).

As an interim measure to help with the problem of LGS size, lenslet
arrays with 2
focal lengths have been installed, giving pixel scales of
0.75\arcsec\,pixel$^{-1}$ and 1.25\arcsec\,pixel$^{-1}$.
The phase error due to read noise \cite{rou94} is
\[
\sigma^2 \ \ = \ \ 
			\frac{\pi^2 \ \sigma_{\rm e}^2 \ N_{\rm s}^4}
					{3 \ n_{\rm ph} \ N_{\rm d}^2}
\]
where $\sigma_{\rm e}$ is the electron read noise per pixel, 
$N_{\rm d}$ is the FWHM of the diffraction limit in pixels, and
$N_{\rm s}$ is the number of pixels used in the centroiding.
Hence increasing the pixel scale gains us a factor of 3 in the read noise
phase error.

Additional errors can arise with a LGS larger than the isoplanatic
patch $\theta_0$ at the wavelength of the science observations,
particularly if the size of the apertures on the wavefront sensor is
well matched to the coherence length $r_0$. 
However, this is not a serious problem for us, since our system is
designed to operate in the near-infrared where $\theta_0$ is relatively
large, typically 7\arcsec\ at K for 1\arcsec\ seeing.

\subsubsection{Brightness}
\label{sec:bright}

The routinely obtainable laser output power is 3.8\,W.
However, this is reduced to typically 2.1--2.3\,W by the time it
reaches the telescope due to the multiple reflections from mirrors
which very quickly become dirtied with dust blown over from the Sahara.
More recently, in mid-1999, we have been achieving output powers of
4.1\,W and, with freshly cleaned mirrors, a launch power of 3.2\,W

The addition of COT to the Rh6G glycol dye solution increases the laser
power by about 10\% by further suppressing the triplet states in the
dye molecules (see Rabien et al. this issue).
However, we have found that this lasts only for short time-scales
of tens of minutes.
A further effect is that the lifetime of the molecules is extended,
although the reason for this is unclear.

If the bandwidth of the laser is very narrow compared to the doppler
width of the sodium in the mesospheric layer so that it effectively
excites only a single velocity group, the magnitude $m_{\rm V}$ of the
LGS can be calculated from $P_{\rm obs}$~(W\,m$^{-2}$), the power
observed,
\[ 
m_{\rm V} \ \ = \ \ -2.5 \ \log \ {
			\left(\frac{P_{\rm obs}}{3.24\times10^{-9}} \right) }
\]
This in turn is derived from the estimate
\[
P_{\rm obs} \ \ = \ \ \frac
		{1.2 \ P_{\rm l} \ T_{\rm atm}^2 \ \sigma \ N_{\rm Na}}
		{4 \ \pi \ H^2}
\]
where $P_{\rm l}$~(W) is the launch power,
$T_{\rm atm}$ the atmospheric transmission, 
and $H$~(m) the distance of the Na layer from the telescope. 
Since typically only 4\% of the laser power is absorbed by the Na at 
zenith (and $<$10\% at large zenith angles), that
fraction is well approximated by $\sigma\,N_{\rm Na}$,
where $\sigma = 8.8 \times 10^{-16}$\,m$^2$ is the homogenously 
broadened absorption cross-section of the
D$_2$ transition, and $N_{\rm Na}$~(m$^{-2}$) the column density.
The factor of 1.2 is due to the marginal direction
dependence of the scattered intensity, due to the interaction of Na
atoms with the circularly polarised laser light.
A further modifier should be included if the intensity of the laser in
the mesosphere is close to the saturation intensity of
6\,W\,m$^{-2}$\,MHz$^{-1}$. 
As the laser intensity increases a larger fraction of the fluorescence
power is wasted as stimulated emission near the laser axis, approaching
half of the total when the laser and saturation intensities are equal.
For a 10\,MHz bandwidth
\[
P_{\rm obs} \ \ = \ \ \frac{P_{\rm obs}} {1 \ + \ 
			\left(\frac{P_{\rm l} \ T_{\rm atm}} 
					{1.1\times10^{-9} \ (H \ \theta_{\rm lgs})^2} \right)}
\]
where $\theta_{\rm lgs}$~(arcsec) is the diameter of the LGS.
For our launch power of 2.3\,W this will only be important if the
spot size is less than about 1\,arcsec.

\begin{figure}[h]
\plotone{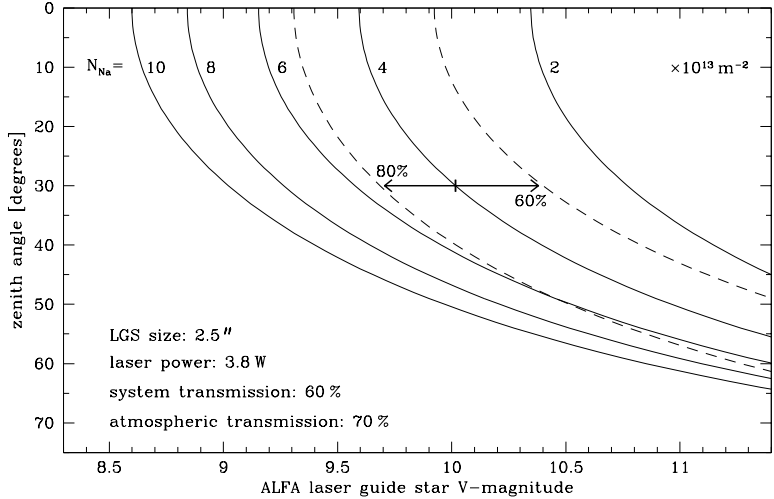}
\caption{Predicted LGS brightness (given as equivalent V-band
magnitude) as a function of zenith angle for various Na column
densities.
Typical values for the laser power, system transmission, and
atmospheric transmission have been used.
These can all drastically affect the magnitude of the LGS, and
as an example, we show how it changes if the atmospheric transmission
increases or decreases by 10\%.}
\label{fig:lgsmag}
\end{figure}

Away from zenith each of $T$, $H$, and $N_{\rm Na}$ 
need to be adjusted appropriately,
resulting in a strong dependence on zenith angle.
Figure~\ref{fig:lgsmag} shows the prediction of
their effect on the LGS magnitude.

During March~1998 the magnitude was measured as m$_{\rm V} = 9.1$  for
an estimated launch power of 2.3\,W and atmospheric transmission of
$T_{\rm atm} = 70$\% each way, from which the Na column density was
derived as $N_{\rm Na} = 6.2 \times 10^{13}$\,m$^{-2}$. 
Typical values measured by from
absorption in the solar spectrum on Kitt Peak and in the stellar
spectra of $\alpha$\,Leo and $\alpha$\,Aql) on Mt Hopkins at a
latitude of 32\deg~N (very similar to the 37\deg~N of Calar Alto) are
2--$6\times10^{13}$\,m$^{-2}$ \cite{ge97,ge98}.
However, comparisons are fraught with difficulties due to the seasonal,
diurnal and even hourly variations in Na column density 
(eg. see \opencite{age99}, and references
therein).

\subsubsection{PSF simulations}
\label{sec:psf}

\begin{table}
\caption{Parameters for simulated K-band PSF in 1\arcsec\ seeing}
\label{tab:psf}
\begin{tabular}{rrrrrr}

\hline

Distance to & residual FWHM & Strehl & FWHM & energy in & 50\% encircled \\

tip-tilt star & image motion & ratio & resolution & 0.20$''$ radius & energy radius \\

[arcsec] & [arcsec] & [\%] & [arcsec] & [\%] & [arcsec] \\

\hline

  0\hspace{12pt} & 0.00 & 25.0 & 0.14\hspace*{1.5mm} & 23 & 0.41 \\
 30\hspace{12pt} & 0.06 & 21.3 & 0.15\hspace*{1.5mm} & 24 & 0.41 \\
 60\hspace{12pt} & 0.10 & 17.2 & 0.17\hspace*{1.5mm} & 25 & 0.41 \\
 90\hspace{12pt} & 0.21 &  9.2 & 0.30\hspace*{1.5mm} & 24 & 0.41 \\
120\hspace{12pt} & 0.46 &  4.1 & 0.66\hspace*{1.5mm} & 16 & 0.43 \\

$\left.\begin{array}{r}$no tip-tilt$\\$correction$\end{array}\right\}$
&    0.72 &   2.9 & 0.90\hspace*{1.5mm} & 16 & 0.47\vspace{1mm}\\

$\left.\begin{array}{r}$seeing-$\\$limited$\end{array}\right\}$
& ---\hspace*{1.5mm} &   2.4 & 1.00\hspace*{1.5mm} & 10 & 0.50\vspace{1mm}\\

$\left.\begin{array}{r}$diffraction-$\\$limited$\end{array}\right\}$
& ---\hspace*{1.5mm} & 100.0 & 0.13$^a$         & 66 & 0.09 \\

\hline

\end{tabular}

$^a$ The commonly used half-width zero intensity is 0.144\arcsec,
slightly {\it better} than the 0.166\arcsec\ for a 3.5\,m mirror with no
central obscuration.

\end{table}

\begin{figure}
\plotone{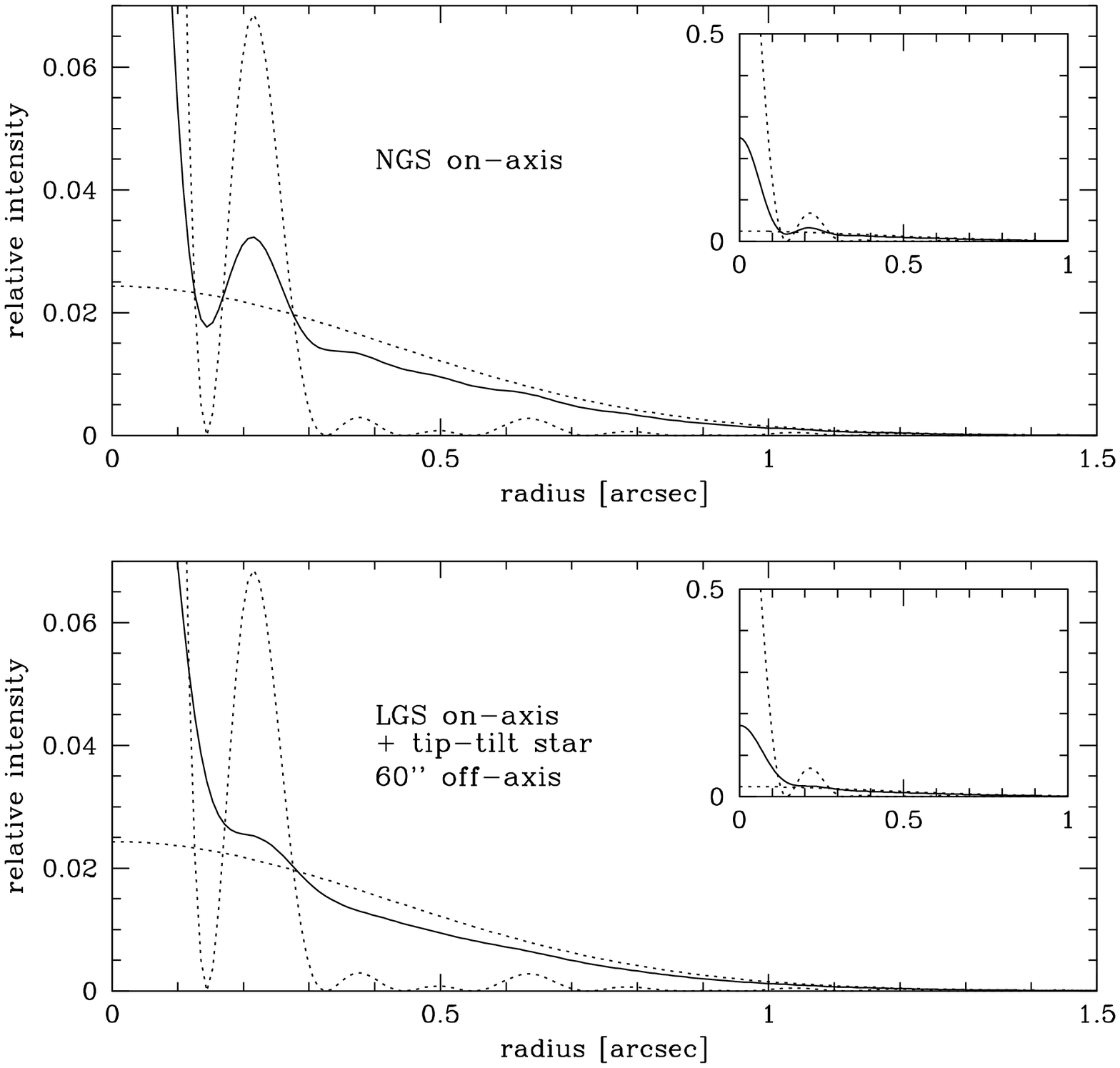}
\caption{Simulated point-spread function profiles which could in
principle be
achieved with ALFA, normalised to a maximum intensity of 1 for the
diffraction limit.
Details of the profiles are shown in the main figures, and the full
profiles are inset.
The dotted lines are the Gaussian seeing profile, and modifed Airy
profile; the solid lines are the simulated PSFs.
Top is that for a natural guide star with a strehl
of 25\%.
Below is that for a laser guide star and a star 60\arcsec\
off-axis for tip-tilt correction. The residual jitter (0.10\arcsec\ FWHM) reduces the
strehl to 17\%,
but makes little difference to the 50\% encircled energy radius.}
\label{fig:psf}
\end{figure}

We have modelled the point-spread functions (PSFs) that might be
expected when ALFA is operating under typical conditions (eg K-band,
1.0\arcsec\ seeing) and summarise the parameters in Table~\ref{tab:psf}.
The atmospheric turbulence parameters give r$_0 = 45$\,cm and a
Greenwood frequency \cite{gre77} of 
$f_{\rm G} = 7$\,Hz, both at 2.2$\mu$m.
For the model we have assumed the seeing profile is Gaussian and the
diffraction limited profile is an Airy function that has been modified
to take into account the 1.37\,m central obscuration in the 3.50\,m
mirror.
The partially corrected PSFs assume the usual model of
a diffraction limited core, and a halo with FWHM similar to the seeing.
When using the $5\times5$ lenslet array only 21 modes are corrected, and
combined with a sampling rate of 100\,Hz (disturbance rejection
bandwidth $\sim$8\,Hz), the residual phase error is about 1\,rad$^2$.
We have included an extra 0.4\,rad$^2$ error to simulate the noise
associated with correcting on a \mv=9.5\,mag LGS.
Thus the Strehl ratio for an on-axis natural guide star (NGS) is about
25\%.
For the LGS we have assumed that the tip-tilt is corrected using an
off-axis star, introducing some residual image motion (given in the
table as FWHM);
but that the star is bright enough that no wavefront
aberrations are introduced by detector noise or bandwidth limitations.
The table clearly shows that much beyond 60\arcsec\ the performance of
the AO is severely degraded.
At 140\arcsec\ the tip-tilt of the star is completely uncorrelated with
that on-axis, because the jitter is the same as that for uncorrected
one-axis tilt \cite{fri65}
\[
\sigma^2 \ = \ 0.18 \  
		\left( \frac{\lambda}{D} \right)^2 \ 
		\left( \frac{D}{r_0} \right)^{5/3}
\]
where $D$~(m) is the telescope diameter, $\lambda$~(m) the observing
wavelength, and $\sigma^2$~(rad$^2$) gives a FWHM jitter of
0.72\arcsec.

The profiles for a tip-tilt star on-axis and 60\arcsec\
off-axis are shown in Figure~\ref{fig:psf}.
The dotted lines are the seeing and diffraction
limited profiles; the solid lines show the simulated PSFs.
When the star is 60\arcsec\ off-axis the residual image motion is
0.1\arcsec\ and reduces the strehl ratio to 17\%.
Whether it matters to the observer depends on the quantities
required:
the 50\% encircled energy radius of 0.41\arcsec\ is the same in both
cases; the fraction of energy within a 0.2$''$ radius actually
increases marginally with an off-axis tip-tilt star (because some energy
from the first Airy ring is shifted into the aperture).
Hence, if getting as much energy as possible into a slit
or simply trying to detect faint objects is the aim, then the
residual tilt is not a problem.
The FWHM resolutions are 0.14\arcsec\ and 0.17\arcsec, both good but
the difference could be important if trying to separate close objects.
On the other hand, the stronger Airy ring in the former case might
offset any gain from a narrower core.
Further analysis of the PSFs obtained with NGS and LGS will be possible
with more observations.

\section{LGS Experiments}
\label{sec:tmr}

During August~1998 a number of experiments were performed at the
Calar Alto observatory in collaboration with the TMR Laser Guide Star
Network\footnote{the Network for Laser
Guide Stars on 8-m class Telescopes operates under the auspices of the
Training and Mobility of Researchers programme of the European Union}, 
during which SCIDAR was carried out on the 1.2\,m to measure
atmospheric turbulence, a MAMA detector on the 2.2\,m measured photon
return from the LGS and Rayleigh cone, and ALFA was operated on the 3.5\,m.
Full results from these will be published elsewhere.
The experiments attempted to tackle a range of issues, and the
rationale for them is outlined briefly below.
More detailed discussion of the methods used and results obtained can
be found in several papers also in this issue (O'Sullivan et al.,
Viard et al., and Esposito et al.).

\subsection{Tip-Tilt from a Laser Beacon}

The tip and tilt modes of a wavefront distortion are responsible for
the bulk of the image degradation.
Each of these terms contributes a phase error of
$0.448(D/r_0)^{5/3}$\,rad$^2$ to the total
$1.030(D/r_0)^{5/3}$\,rad$^2$ \cite{nol76}, and unless these can be
corrected effectively then 
any image improvement from higher order terms will be negligeable.
Although this means that image resolution can often be enhanced significantly
simply by use of a fast guider, it also has implications for laser guide
stars because the up-link and down-link tip-tilt components cannot be
separated.
Thus a laser beacon cannot easily be used to measure tip-tilt modes,
with the result that sky coverage is limited by the availability of
stars for fast guiding.
There are two alternative proposals to circumvent this difficulty.
One involves the polychromatic LGS concept \cite{foy95} in which the
4P$_{3/2}$ level 
of Na is excited, and the ensuing radiative decay results in a line
spectrum from 0.33\,nm to 2.207\um. 
The wavelength dependence of the refractive index of air allows the
tip-tilt component to be determined.
Another method is to use auxiliary telescopes \cite{rag95} from which 
the up-link and down-link motions of the laser and beacon are
uncorrelated.
The uncorrelated tip-tilt motions of the beacon and a natural guide
star are due to the up-link tip-tilt of the laser alone.
The advantage is that there is a much greater chance of finding a
natural star in the same isoplanatic patch as the laser, since it now
appears highly elongated.

In collaboration with the TMR LGS network we attempted the first measurement
of tip-tilt from a laser guide star, using the 2.2-m to view the LGS
off-axis.
Initial analysis of a small part of the data indicate that
sufficiently separated sections of the extended laser plume
show uncorrelated motions, but the same differential motion with
respect to nearby natural stars, which is the up-link motion of the laser.
By removing the jitter measured below the laser launch telescope,
it is possible to reconstruct the motion due to atmospheric tip-tilt
alone.
A second experiment to compare the tip-tilt measured in this way to
that observed by ALFA from an on-axis natural star proved to be more
complex and further observations will be needed.

\subsection{Site Characteristics}
\label{sec:sitechar}

It is very important to have some knowledge of the atmospheric
characteristics of the site if successful and optimal adaptive optics
correction is to be achieved.
To this end, a number of experiments have been monitoring the profile
of the Na layer, atmospheric turbulence, and the performance of ALFA.

\subsubsection{Sodium layer profile}

Viewed from the 2.2\,m telescope, the LGS plume appears to be about
100\,arcsec\ long, and hence can be used to look at the vertical
profile of the sodium layer with a resolution better than 250\,m.
Previous observations during the autumn of 1997 have shown that over
the time-scale of 1\,hour the profile of the Na layer can change
significantly.
It is crucial to know the time-scale on which these structures appear
and disappear, and the scale of the differences.
Narrow sporadic layers have been observed, and whether they are
actually high-altitude noctilucent clouds or real changes in Na
abundance can be established by tuning the
laser on and off the Na line.
The main consequences of such variability are the effect on the
brightness and centroid height of the LGS.
Simulations suggest that for a well-focussed spot it will be necessary
to refocus the wavefront sensor, and perhaps also the launch telescope,
if the effective altitude changes by more than 1\,km.
The height and distribution of the Na at various times during the night
have therefore been monitored \cite{osu99}.

Future possibilities for more permanent measurement of the sodium, in
order to gain detailed statistics of the site, include using ALFA as a
LIDAR station.
The laser will be amplitude-modulated to give micro-second pulses,
allowing the time of flight of the photons to be measured, and hence a
vertical density profile of the sodium layer to be calcuated.
The backscattered light will collected by the telescope will be fed
into an APD.
A psedo-random sequence of on/off pulses in a non-return to zero format
will allow a high signal-to-noise to be reached in a short time.
The system has the great advantage that the Na profile can be measured in
a few seconds, at almost any time and with little extra effort required.

A different monitoring experiment has begun to determine the effects
and seriousness of light pollution from the laser due to Rayleigh and
Mie backscattering \cite{del99}.
This is an extremely important issue for crowded observatory sites, and
preliminary work at Calar Alto suggests that multiple backscattering
may exceed the ambient background out to distances of $\sim$1\deg\ from
the laser beam.

\subsubsection{Atmospheric turbulence profile}

The effectiveness of any given AO configuration (sampling frequency,
number of modes corrected, etc.) depends on the prevailing atmospheric
conditions at the time.
For the example in Section~\ref{sec:psf}, the 25\%  Strehl we expect with
1\arcsec\ seeing would vary between 19--33\% with seeing in the range
0.8--1.2\arcsec. 
A more practical approach is to measure the average turbulence and
velocity profiles at a particular site and time of year to gain an
understanding of the isoplanicity and time-scale for temporal variation.
The SCIDAR (SCIntillation Detection and Ranging) technique involves
autocorrelating the scintillation patterns due to
a binary star in the telescope pupil plane in order to recover the
spatial correlation and hence 
refractive index structure constant C$_n^2$(h) as a function of height.
The method also allows the wind speed and direction of the dominant
layers to be established by correlation of frames with different
temporal separations.
Data from one night in 1997 suggest $r_0 = 18$\,cm (at $\lambda =
500$\,nm), an  unexpectedly favourable result given the median seeing;
more measurements are needed to determine the typical characteristics.
Further results were obtained during August 1998, and will be analysed
in conjunction with {\it simultaneous} closed loop gradients from ALFA
and the resulting PSF images.
We can then, for example, check the calibration of SCIDAR and ALFA by
comparing the value of $r_0$ derived from each of the above
instruments through roughly the same column of atmosphere; 
from SCIDAR by knowledge of the
turbulence layers and from ALFA via the modal power spectral density.
Similarly the wind velocities and temporal power spectral density can
be compared via the Greenwood frequency.

\section{Observational Results}
\label{sec:res}

For an AO system to work effectively with a LGS as the wavefront
reference, it must first be tested using natural guide stars.
During August 1998 this was done extensively.
On bright stars (V$\simlt$8) it is now possible, even in relatively
poor seeing conditions, to obtain diffraction limited images
routinely with Strehl ratios at least 40\%.
On stars as faint as V=11\,mag it is still possible to achieve Strehl
ratios of 10--15\%, with a resolutions of around 0.25$''$.
Currently the faintest object on which we have locked is NGC\,1068 which
has an unresolved core fainter than \mv$\sim$13\,mag.

These results indicate that the AO is at a stage where routine
correction using a LGS should be possible, and so the system is now
open to guest observers at Calar Alto on a shared-risk basis.
However, the problems discussed previously mean that LGS observations
are still far from routine.
In the rest of this Section we discuss 2 sets of observations.
The first is of the field around
a Herbig Ae/Be star which show that ALFA is effective on relatively
faint stars, and that in the partial correction mode the isoplanatic
patch is large.
The second is of 2 galaxy clusters, and include the first image
improvement on another galaxy using the laser guide star.
Details of other scientific results, including the first diffraction
limited integral field spectroscopy, can be found in
\inlinecite{dav98e} and \inlinecite{dav99a}.

\subsection{The Field around the Herbig Ae/Be star BD\,+40\deg\,4124}
\label{sec:bd40}

\begin{figure}
\plotone{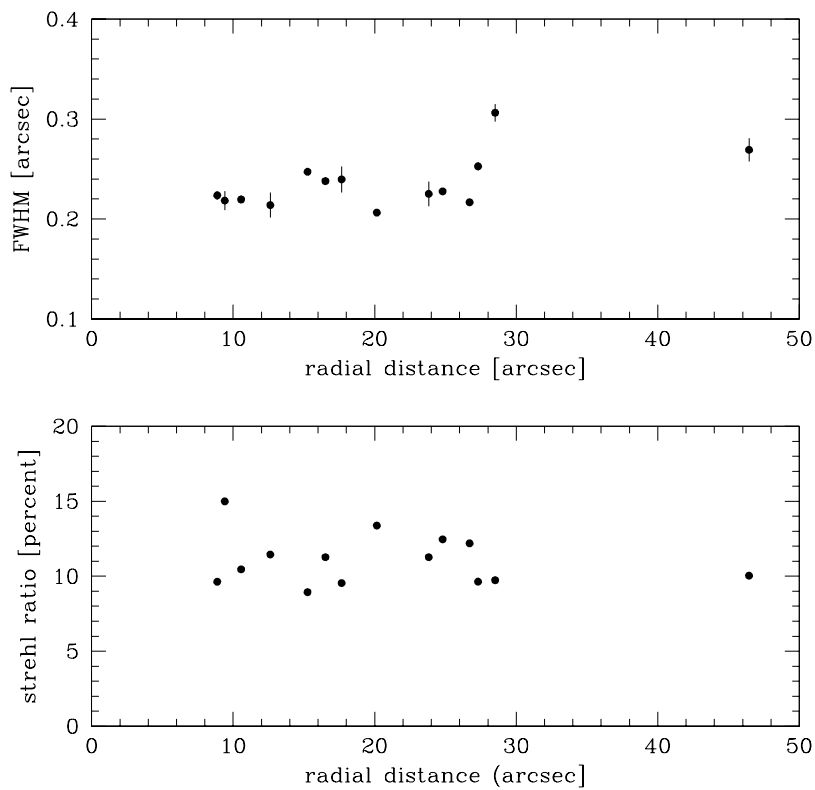}
\caption{Strehl ratio and FWHM of 14 stars in the field around
BD\,+40\deg\,4124, plotted against distance from this star.
The advantage of working in the partial correction regime is clear,
since the effects of anisoplanaticism are much reduced:
only beyond 30$''$ is the a small reduction in performance.}
\label{fig:bd40.strehl-fwhm}
\end{figure}

During August~1998 the Herbig Ae/Be star BD\,+40\deg\,4124 was observed
with ALFA, using the 3D integral field spectrometer as well as
Omega-cass;
here we discuss issues relevant to the AO performance.
For the wide field K-band image with Omega-cass, BD\,+40\deg\,4124
(V=10.6, K=5.6) was centered in the wavefront sensor,
allowing correction of 18 modes with a frame rate of 75\,Hz.
Although this star was saturated the Strehl ratios and FWHMs of 14
others in the $80''\times80''$ field were measured, and these are
plotted against their radial 
distance in Figure~\ref{fig:bd40.strehl-fwhm}.
The pixel size of 0.08$''$ means that the Strehls could be
underestimated by as much as 4\%, and the scatter may be partly due to
this.
The FWHMs were estimated using only vertical and horizontal cuts (the
`errorbars' are lines joining these points) since models suggest that
the pixel size is already limiting the resolution.
The figure indicates that out to radii of at least 30$''$ there is very
little degradation in performance since the wavefront error due to
anisoplanaticism is small compared to the total error;
and beyond this as far as can be measured, the correction is only
marginally worse.

\begin{figure}
\epsscale{0.5}
\plottwo{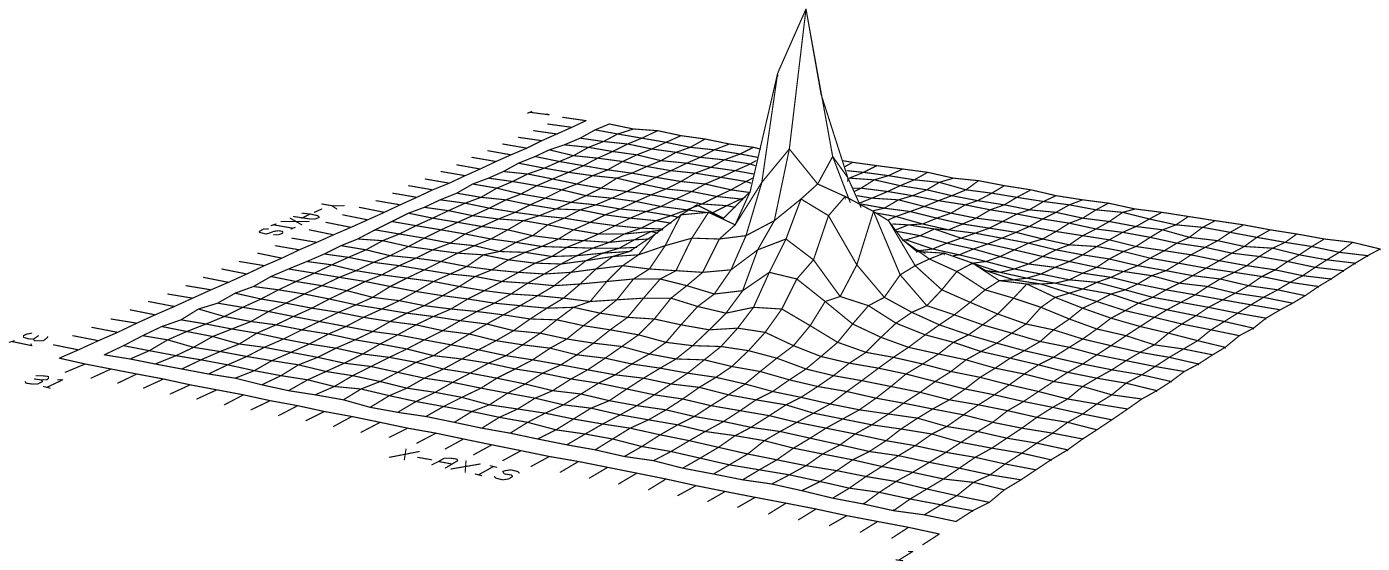}{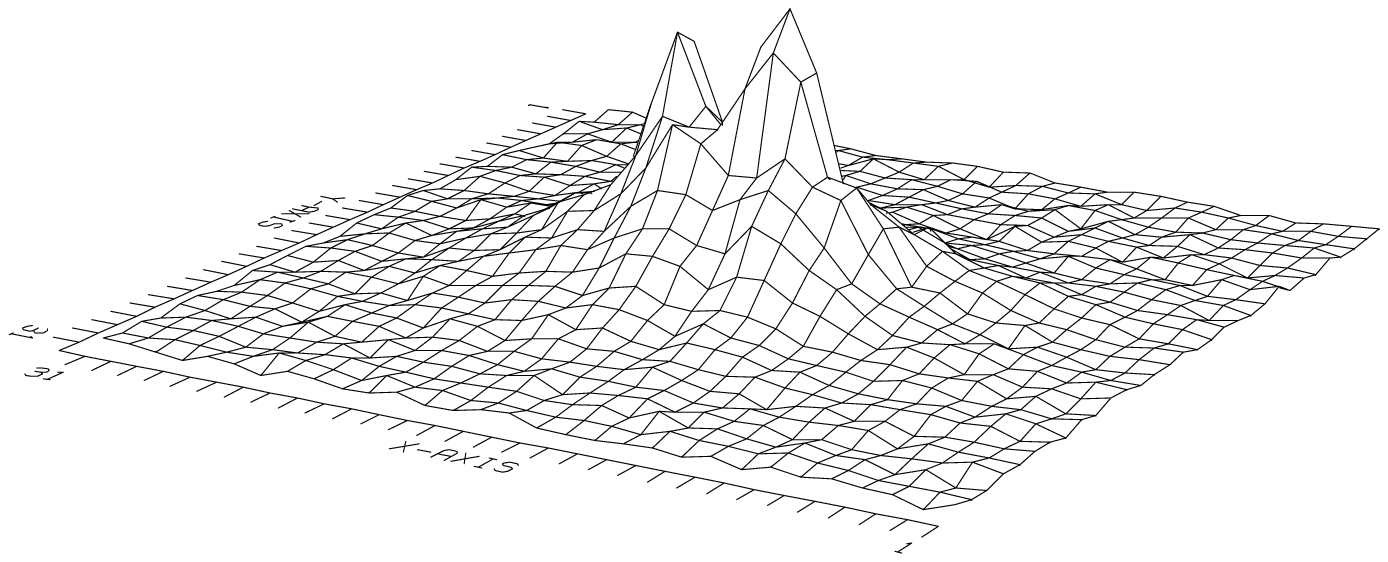}
\caption{K-band surface plots of star 14 (left) and the binary system
 (right), both approximately 20$''$ from BD\,+40\deg\,4124.
The pixels are 0.08$''$ wide and the resolution is
 sufficient to separate stars 0.32$''$
 apart, as shown by the previously unknown binary system here.
The profiles of all the stars show that the PSF changes very little
 even out to distances of 50$''$.}
\label{fig:bd40.psf}
\end{figure}

Figure~\ref{fig:bd40.psf} shows that the correction achieved here is
useful: a star 20$''$ away clearly has a single peak (as do all the
single stars in the field), while another has
a double peak, identifying it unambiguously as a previously unknown
binary system with a separation of 0.32$''$.
Importantly, examination of the profiles of all the sources indicates
that the speckle pattern in the halo appears to be stable over the
whole field.
This means that deconvolution should work well (yielding an effective
resolution rather better than 0.2$''$) and that the choice of
PSF, at least in this case, is not critical.
A further result from this image was that a star in the pair
V1318\,Cyg (40$''$ East of BD\,+40\deg\,4124) was resolved to be
$0.4''\times0.3''$.
Since JHK colours also indicate that this pair are both very red, we
interpret it as a dust shell around the star, with a deconvolved size of
$300\times170$\,AU (assuming a distance to the stellar aggregate of
1\,kpc, \opencite{hil95}).

\subsection{Galaxies in the Abell 1367 and 262 Clusters}

Between December~1997 and August~1998 we observed 25 galaxies in
the Abell\,1367 and Abell\,262 clusters, including two barred spiral
galaxies UGC\,1347 and UGC\,1344 
(full results and analysis will be given in Eckart et al. 1999, in
preparation).
 
\begin{figure}
\epsscale{1}
\plotone{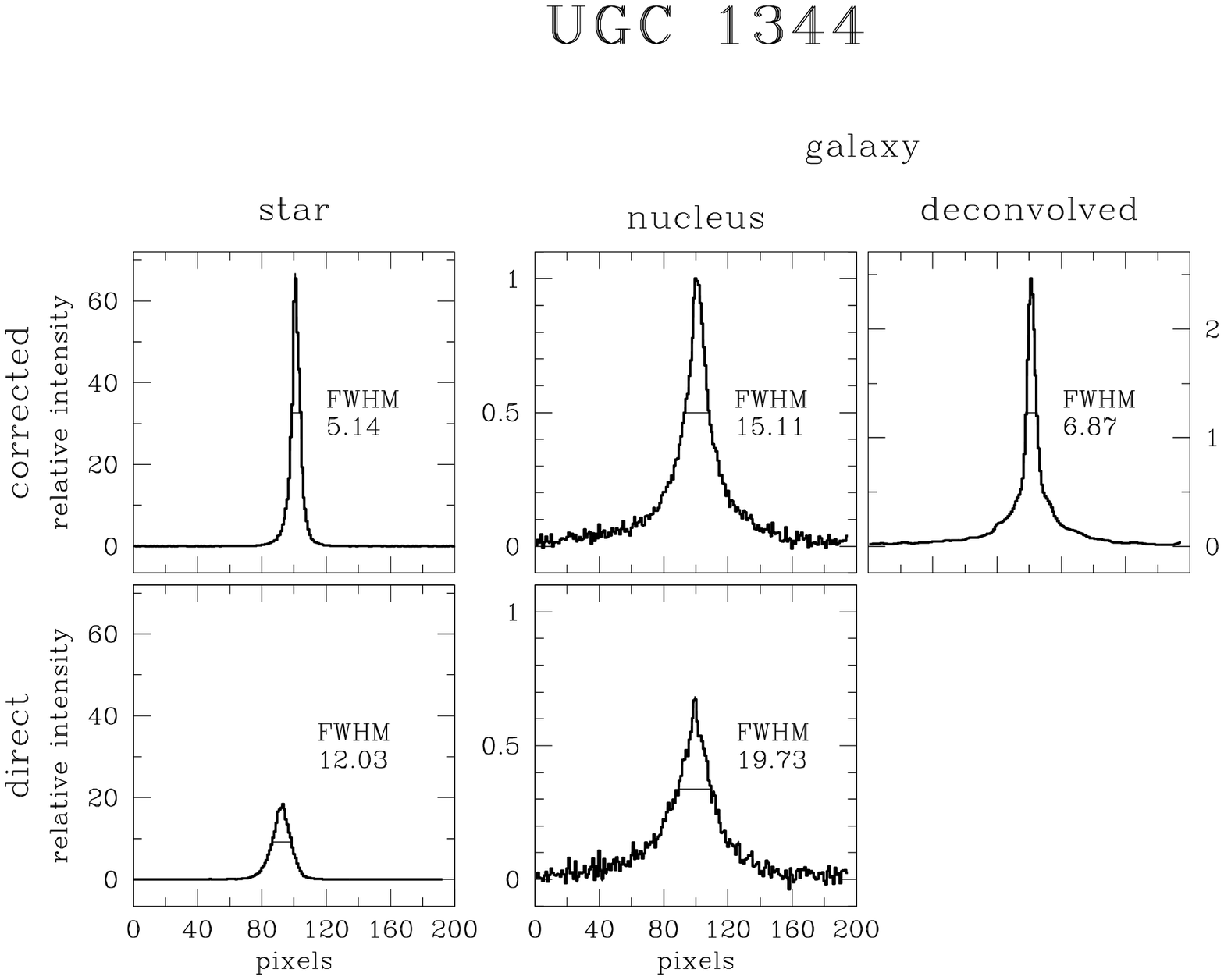}
\caption{Profiles of the star (left) and nucleus of UGC\,1344 (right)
 in both open loop (lower) and closed loop (upper);
a deconvolved closed loop profile is also shown (far right).
FWHM are given in pixels, with a scale of 0.08\arcsec.
The peak intensity of the star increases by a factor of 3, while the
galaxy nucleus increases by a factor of only 1.4 suggesting that there
is a narrow core component and a wdier bulge component.}
\label{fig:ugc1344}
\end{figure}

For UGC\,1344 we used a nearby (27$''$ separation) natural guide star with
V=11.0\,mag, correcting 7 modes at a sampling rate of 150\,Hz,
achieving a disturbance rejection bandwidth of about 13\,Hz.
Cuts of the profiles through the star and galaxy, for both open and
closed loop as well as deconvolved are shown in
Figure~\ref{fig:ugc1344}.
The peak intensity of the star increased by a factor of 3, and the FWHM
improved from 1.0$''$ to 0.4$''$.
For the galaxy, the peak increased by only 1.4, although
the discussion for BD\,+40\deg\,4124 indicates that the galaxy is
well within the isoplanatic patch.
The deconvolved image begins to indicate the reason: there is a narrow
unresolved core and a wider bulge component so the 
increase in peak intensity will be that due to the core only.
The observations can be explained if the flux in the bulge is 3 times
that in the core.
The presence of a core suggests a recent localised burst of star
formation in the galaxy's nucleus.

\begin{figure}
\plotone{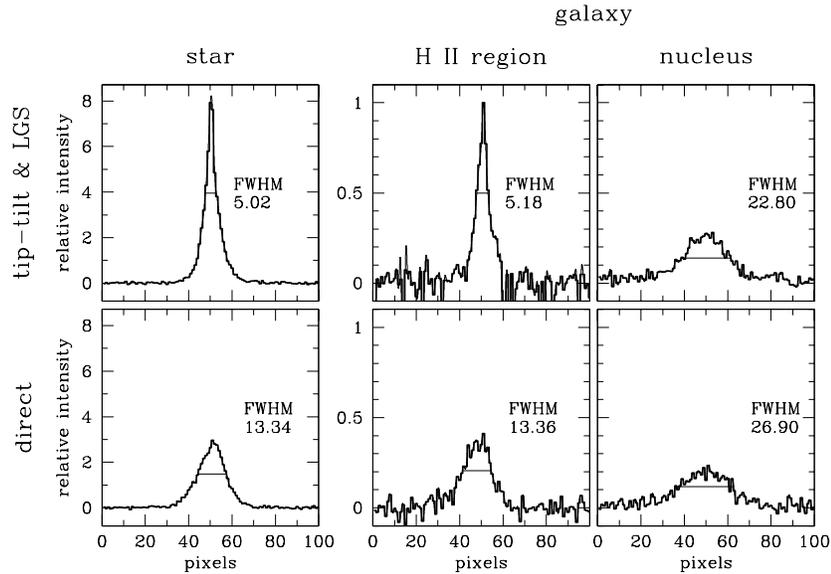}
\caption{Profiles of the star (left), and the compact \hiiregion\ and
nucleus of UGC\,1347 (right) in both open loop (lower) and closed loop
(upper).
FWHM are given in pixels, with a scale of 0.08\arcsec.
The peak intensity of the star increases by a factor of 2.5, as does
the unresolved \hiiregion, while there is almost no enhancement in the
nucleus showing it is completely resolved.
The laser was used to correct the high orders in these images, while
tip-tilt was determined from the star.
Tip-tilt alone produced little improvementin image quality.}
\label{fig:ugc1347}
\end{figure}

For UGC\,1347 we corrected tip-tilt using a V=11.8\,mag star 41$''$
away, and pointed the laser guide star midway between this and the
galaxy.
High order corrections were achieved using the LGS, also with 6
subapertures but at a lower frame rate of 50\,Hz (rejection
bandwidth of only 4\,Hz).
Figure~\ref{fig:ugc1347} shows for the star an increase in peak
intensity of 2.5, and improvement in resolution from 1.1$''$ to
0.4$''$.
Exactly similar enhancements are seen in the compact \hiiregion\ of the
galaxy 11\arcsec\ from the nucleus, but almost no change is seen in
the nucleus itself.
All the evidence for recent star formation in this galaxy can be
accounted for by the \hiiregion\ alone.

Both these galaxies have projected positions close to the centre of the
Abell\,262 cluster, and both have large velocities with respect to the
cluster average.
However, only UGC\,1344 is H{\sc i} deficient \cite{gio85}.
H{\sc i} deficiency strongly correlates with distance from the cluster
centre, the central region being the zone of depletion.
The interpretation is that gas is stripped away as the galaxy crosses
the crowded central region, giving rise to the hot intracluster gas
observed in X-rays.
Our results for the whole sample suggest also that star formation is induced
as a galaxy crosses the centre of the cluster.
In this interpretation, the H{\sc i} deficient galaxy UGC\,1344 has
just passed the centre so that it has lost its gas and undergone a
burst of star formation;
UGC\,1347 is approaching the centre and has not yet been depleted of
its H{\sc i} nor had nuclear star formation triggered.

\subsection{Achieving High Strehls with a LGS}

These results show that the potential for laser guide stars is there,
but the gains achieved so far have been marginal.
Following optimisation of the launched laser beam and also of the
wavefront sensing hard- and software, we have reached a Strehl ratio of
23\% in the K-band using the LGS to correct high-orders.

\begin{figure}
\centerline{\psfig{file=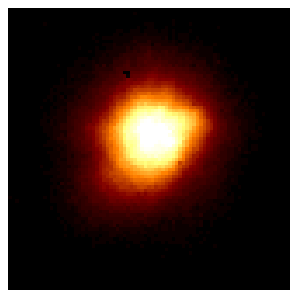,width=25mm}\hspace{2mm}\psfig{file=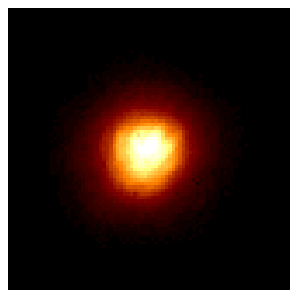,width=25mm}\hspace{2mm}\psfig{file=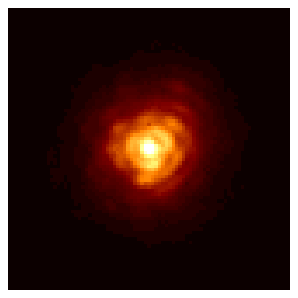,width=25mm}\hspace{2mm}\psfig{file=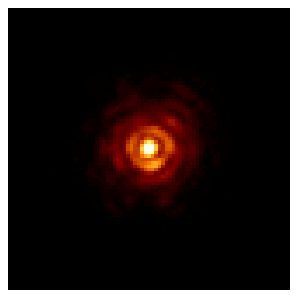,width=25mm}}
\centerline{\hspace{-2mm}\psfig{file=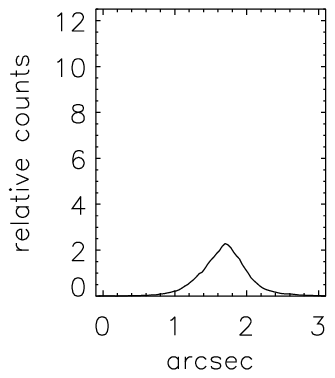,width=30mm}\hspace{-2mm}\psfig{file=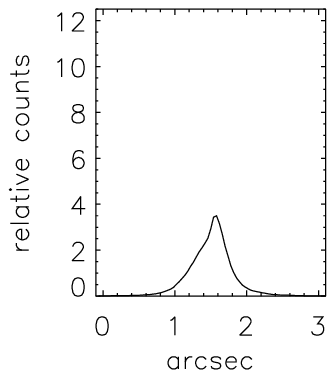,width=30mm}\hspace{-2mm}\psfig{file=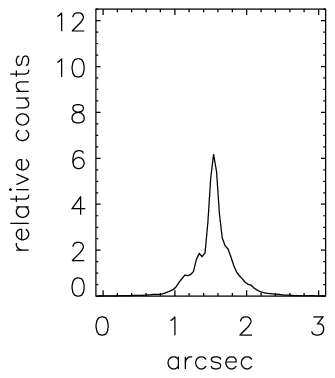,width=30mm}\hspace{-2mm}\psfig{file=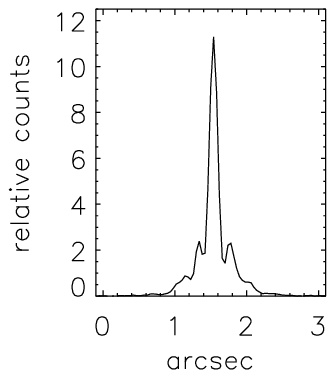,width=30mm}}
\caption{Images of and cross-sections through the star SAO\,68075 during: 
open-loop observing with seeing of 0.65\arcsec\ (far left), 
tip-tilt correction on the star (centre left), 
high-order correction on the LGS (centre right), 
and the best 10 seconds of high-order correction (far right).}
\label{fig:beststrehl}
\end{figure}

For these results, shown in Figure~\ref{fig:beststrehl},  the laser was
pointed on-axis and
imaged on the WFS through the 5x5 lenslet array at a frame
rate of 75\,Hz. This allowed 18 high order Karhunen-Loeve modes to be
corrected with a bandwidth of around 12\,Hz. 
A natural star (SAO\,68075), 10\arcsec\ off-axis to avoid any of its
light reaching the WFS, was used to correct tip-tilt motion at 65\,Hz. 
A series of 40 consecutive images were taken of the star in the K-band.
Each integration was 5\,sec, and the only processing performed was
sky-subtraction and bad pixel removal.
In particular, no shift-and-add or deconvolution techniques were
applied. 
Frames were added together directly to produce long exposure images.

During the observations the K-band seeing was 0.65\arcsec, a remarkably
good value.
A 50\,sec image with the tip-tilt corrected on the star shows the
resolution with this alone was improved to 0.49\arcsec.
Adding together all 40 images taken while high-orders were corrected
with the LGS gave an effective exposure time of 200\,sec, during which
the FWEHM was reduced to 0.19\arcsec, and a Strehl ratio of 12.8\% was
reached.
During the best 10\,sec a Strehl of 23\% was achieved and the image is
diffraction limited.
Probably the performance is variable because of the jitter of the LGS
on the WFS, which due to techincal problems has not yet been corrected.
This is entirely atmospheric, due to the different up and down tilt
components, and means that the centre of the LGS spots must often be
determined while close to the boundary of the centroiding box.
Additionally, there may be non-linear effects in the modal
reconstruction due to the large offsets from the zero-point calibration
positions.
Nevertheless, the superb performance achieved here shows the exciting
future potential of adaptive optics with laser guide stars.

%
%

%

\begin{acknowledgements}
The MPIA/MPE team thanks the Calar Alto staff for their help and
hospitality, and N. Wilnhammer for technical support.
RID acknowledges the support of the TMR (Training and Mobility of
Researchers) programme as part of the European Network for Laser Guide
Stars on 8-m Class Telescopes.
Thanks are also due to the referee for a number of pertinent comments
from which the manuscript has benefited.
\end{acknowledgements}

%
%

%

%

\bibliographystyle{klunamed}
\bibliography{/afs/mpa/home/davies/reference}

%
%
%

\end{article}

\end{document}